\documentstyle[aps,twocolumn]{revtex}
\begin{document}
\draft
\title{Hole depletion and localization due to 
disorder in insulating $\bf PrBa_2Cu_3O_{7-\delta}$: 
a Compton scattering study\\}
\author{Abhay Shukla$\rm ^1$, Bernardo Barbiellini$\rm ^2$, 
Andreas Erb$\rm ^3$, Alfred Manuel$\rm ^3$\\
Thomas Buslaps$\rm ^1$, Veijo Honkim\"{a}ki$\rm ^1$,
and Pekka Suortti$\rm ^1$}
\address{$\rm ^1$European Synchrotron
Radiation Facility, BP 220, F-38043 Grenoble, France\\
$\rm ^2$Northeastern University, Department of Physics,
Boston, MA 02115 USA
\\
$\rm ^3$DPMC,Universit\'e de Gen\^^ eve, 
24 quai E. Ansermet, CH-1211 Geneva 4,
Switzerland}
\date{}
 
\maketitle
 
\begin{abstract}
The (mostly) insulating behaviour of $\rm PrBa_2Cu_3O_{7-\delta}$ 
is still unexplained and even more interesting since the 
occasional appearance of superconductivity in this material. Since 
$\rm YBa_2Cu_3O_{7-\delta}$ is nominally iso-structural and always 
superconducting, we have measured the electron momentum density in 
these materials. We find that they differ in a striking way, the
wavefunction coherence length in $\rm PrBa_2Cu_3O_{7-\delta}$ being
strongly suppressed. We conclude that Pr on Ba-site substitution 
disorder is responsible for the metal-insulator transition.
Preliminary efforts at growth with a method to prevent
disorder yield 90K superconducting 
$\rm PrBa_2Cu_3O_{7-\delta}$ crystallites.
\end{abstract}
\pacs{PACS numbers: 74.72Jt, 74.72Bk, 74.62Dh, 78.70Ck}
\narrowtext

The peculiar case of $\rm PrBa_2Cu_3O_{7-\delta}$ stands out in 
the $\rm RBa_2Cu_3O_{7-\delta}$ family ($\rm R$ for rare 
earth or $\rm Y$) because it is the only easily fabricated 
member which until recently was consistently
insulating in the normal state {\em and} non-superconducting 
(for a review see \cite{radousky,booth}).
Superconducting samples have since been fabricated 
\cite{blackstead,zou}, some of which show metallic behaviour 
in the normal state, akin to that seen in
$\rm YBa_2Cu_3O_{7-\delta}$. The non-superconducting behaviour 
has been attributed to various causes: 
(1) +4 valence for Pr leading to hole filling in the Cu-O layers,
(2) hybridization between the Pr 4f and O 2p orbitals 
\cite{fehrenrice}, leading to hole localization 
in these layers, (3) a chain-based model 
for superconductivity and magnetic 
pair-breaking induced by Pr on the Ba site \cite{blackstead}.
The first possibility has 
been invalidated on the grounds that Pr on the rare-earth
site is essentially trivalent \cite{soderholm,jostarndt} 
while the last is unlikely if only because it fails to explain why 
non-superconducting $\rm PrBa_2Cu_3O_{7-\delta}$ 
should be insulating in the normal state. 

The Fehrenbacher-Rice model of hole-localization 
in its various forms\cite{fehrenrice,liechmazin} remains a 
possible explanation, and to get further insight into this matter 
we have measured the electron momentum density in both 
$\rm YBa_2Cu_3O_{7-\delta}$ and $\rm PrBa_2Cu_3O_{7-\delta}$ 
using Compton scattering of high energy X-rays.
This is the first high statistics and high resolution
measurement in these materials, which are difficult 
candidates for such experiments given the problems related
to fluorescence background from heavy atoms and low 
count-rates due to absorption. These problems have
been overcome thanks to high energy synchrotron 
radiation and dedicated instrumentation 
at ID15B at the ESRF \cite{manninen,suortti}.
Our measurements show that 
the experimental momentum density is similar 
to that given by LDA band-structure calculations for 
$\rm YBa_2Cu_3O_{7-\delta}$. However, in 
$\rm PrBa_2Cu_3O_{7-\delta}$, a striking difference 
is seen between measurement and calculation.
Simple models show that Pr on Ba site defects
(labelled $\rm Pr_{Ba}$ hence-forward) and a lack of hole
doping in the Cu-O planes probably lead to this difference.
These facts are 
interpreted in the light of other experiments 
\cite{soderholm,jostarndt,hoffmann,%
shukla,takenaka,suzuki,nutley,kinoshita,erb2,guillaume,cava,%
jorgensen,reyes,matsuda,neumeier,guan} 
in the $\rm Y_{1-x}Pr_{x}Ba_2Cu_3O_{7-\delta}$ system.

The experiments were performed on crystals grown by the 
flux method in the non reactive crucible material $\rm BaZrO_3$,
known to give uncontaminated, high-purity samples \cite{erb}. 
The $\rm YBa_2Cu_3O_{7-\delta}$ crystals were 
annealed under 1 bar of oxygen at $\rm 530^{\circ}C$ 
for 170 hours yielding a $\rm T_c$ of 91.3 K and an oxygen 
content of 6.89 \cite{lindemer1}. The $\rm PrBa_2Cu_3O_{7-\delta}$ 
crystals were annealed under 1 bar of oxygen at 
$\rm 500^{\circ}C$ for 200 hours yielding an oxygen content 
of 6.92 \cite{lindemer2}, and were 
insulating and non-superconducting. 
Both exhibited uniform twins in the {\em a-b} plane, 
as is usually observed after such a treatment. 
These twinned crystals were used for the experiment.

Compton scattering, or inelastic scattering with very 
high momentum and energy transfer is a probe of the 
ground state, one-electron properties of the system 
\cite{malcolm,ericphil}. 
The measured quantity $J(p_{\rm z})$, the Compton profile,
is a projection onto the z-direction (parallel to the
scattering vector) of the three dimensional 
electronic momentum density $\rm \rho({\bf p})$:
$ J(p_{\rm z}) =\int_{-\infty}^{\infty}
\rho({\bf p}) dp_{\rm x} dp_{\rm y}$.
Compton profiles were measured using the high-resolution 
scanning spectrometer at ID15B.
A cylindrically bent focusing Si(311) Bragg monochromator
provided an incident energy of 55.91 keV for 
$\rm YBa_2Cu_3O_{7-\delta}$ and 57.78 keV
for $\rm PrBa_2Cu_3O_{7-\delta}$ respectively.
The radiation scattered at an angle of 159 degrees was 
analyzed by a cylindrically bent Ge(440) analyzer. 
For both compounds three twinned single 
crystals were stacked to use the full beam size of 0.2x5 mm. 
Compton profiles were measured for the [100/010] and 
[110] directions, in symmetric reflection 
geometry at room temperature with high statistics 
($\rm 5.10^6$ counts in a 0.05 a.u.\  bin at $\rm p_z=0$; 
1 atomic unit (a.u.) of momentum $\rm \cong$ 1.89 
$\rm \AA{}^{-1}$ ).
The effective resolution was 0.15 a.u. (full width at half 
maximum). The incident beam was monitored using a Si PIN diode 
and this was used to normalize the scanned profiles. 
The raw profiles were further corrected
for geometrical effects, for absorption in air and in 
the sample and for the analyzer reflectivity. The background was 
measured separately and subtracted. Finally the profiles were 
symmetrized and normalized appropriately.

In the following we shall discuss the anisotropy of the 
Compton profiles, that is, the difference between Compton 
profiles measured in two crystallographic directions 
($\rm J_{[100/010]}-J_{[110]}$).  This procedure has the
advantage of getting rid of the isotropic contribution 
of the core electrons as well as residual background.
Since both measured directions are in the {\em a-b} plane,
the structure in the anisotropy originates
essentially in this plane, but includes contributions from all
entities of the unit cell, the Cu-O chains, the Ba-O and Y/Pr 
planes as well as the Cu-O layers. However, from earlier momentum 
density measurements \cite{hoffmann} as well as other experiments 
\cite{shukla,takenaka,suzuki} it has been
established that the Cu-O chains are identical in both materials, 
being locally metallic with a corresponding Fermi surface 
(even in $\rm PrBa_2Cu_3O_{7-\delta}$). Thus,
any differences between the two originate in the remaining 
entities, the Cu-O layers or the Ba-O and Y or Pr planes. 

For comparison with experiment we have computed theoretical 
Compton profiles based on LDA band-structure. A FLAPW 
\cite{massidda} calculation was performed for 
$\rm YBa_2Cu_3O_{7-\delta}$. 
For $\rm PrBa_2Cu_3O_{7-\delta}$, earlier studies 
have shown that two of the three Pr-f electrons can actually
be considered as core states \cite{biagini} 
since treating them as band electrons does not modify the 
density of states significantly \cite{AmbroschDraxl}, implying
that theory predicts little change with respect to 
$\rm YBa_2Cu_3O_{7-\delta}$.
The expected difference in the anisotropy in going 
from $\rm YBa_2Cu_3O_{7-\delta}$ to $\rm PrBa_2Cu_3O_{7-\delta}$
is due to the hybridization of Pr-f electrons with
O-2p electrons which even in LDA calculations produces
some hole depletion in the Cu-O layers similar to the
Fehrenbacher-Rice model\cite{singh}.
However, LDA calculations are known 
to work reasonably well only for the metallic phase and not 
for undoped insulating Cu-O layers in these materials.
To estimate this difference due to hybridization we used the
LMTO\cite{barbiellini} method to calculate the anisotropy
for $\rm PrBa_2Cu_3O_{7-\delta}$ (with the
Pr-f electrons considered as valence electrons), from which
the LMTO anisotropy for $\rm YBa_2Cu_3O_{7-\delta}$ was 
subtracted. This difference should give us an upper bound
on the expected change since LDA is known to overestimate 
the effect of hybridization of localized f-orbitals
(see for example \cite{vasu}).
This difference was then added to the FLAPW anisotropy for
$\rm YBa_2Cu_3O_{7-\delta}$ to give an estimation of the
theoretical anisotropy for $\rm PrBa_2Cu_3O_{7-\delta}$.
Liechtenstein and Mazin \cite{liechmazin} have included 
Coulomb correlations in their LDA+U, LMTO calculations for 
$\rm RBa_2Cu_3O_{7-\delta}$ and find that in R=Pr, 
with respect to R=Y, an additional band crosses the
Fermi level trapping holes and changing their nature, 
but with negligeable impact on the Compton profile
according to our estimations.

The anisotropy is shown as a percentage of the peak value of the 
Compton profile. The amplitude of the anisotropy in the
theory is about 40\% larger than in the experiment.
This tendency is commonly observed, even in simple metals,
and attributed to correlation effects under-estimated by LDA 
\cite{loupias,schneider}. However the shape of the anisotropy, 
which is what we use in our analysis,
is generally very well described by LDA, even in highly 
correlated electron systems containing f electrons, 
such as $\rm CeCu_2Si_2$ \cite{vasu}.
In the figures shown we have scaled the theoretical 
anisotropy to the experimental 
one by a factor of 1.4 to ease comparison. Fig.\ \ref{fig1} shows 
the measured and calculated (solid line) anisotropy
for $\rm YBa_2Cu_3O_{7-\delta}$. The two are remarkably alike,
showing that the overall description of the electronic structure
is satisfactory. 
We also show the calculated 
(as detailed above) anisotropy for $\rm PrBa_2Cu_3O_{7-\delta}$
(dashed line), which, despite Pr-O hybridization, shows
little change with respect to the $\rm YBa_2Cu_3O_{7-\delta}$ 
calculation.
Fig.\ \ref{fig2} shows the measured anisotropy for  
$\rm PrBa_2Cu_3O_{7-\delta}$ which is strikingly different than
the one predicted by theory.

When $\rm PrBa_2Cu_3O_{7-\delta}$ is synthesized it is known that
Ba-rich or Pr-rich, off-stoichiometric phases 
($\rm Pr_{1\pm x}Ba_{2\mp x}Cu_3O_{7-\delta}$) may be produced.
There is however substantial evidence that even in globally 
stoichiometric samples substitution disorder 
concerning the Pr and Ba sites 
(spinodal decomposition for example) exists locally 
\cite{blackstead,nutley,kinoshita,erb2}, especially if growth
is followed by slow-cooling.
Such disorder, due to the large size of Pr ions has also been
observed in the case of similar sized Nd \cite{erb2} 
and La \cite{blackstead,guillaume} ions.
The amount of this substitution has been estimated to be 
a maximum of about 8\% \cite{booth,nutley}. If such disorder 
were present, one effect would be to diminish the contribution
of the Pr and Ba-O planes to the total anisotropy due to
the absence of long-range order in these planes.
Since accounting for such disorder in the theory is not trivial, 
for qualitative analysis we have used a simple model 
within the LMTO method, where we have neglected 
the contribution of Pr and Ba-O planes 
to the total anisotropy (solid line in Fig.\ \ref{fig2}). 
In the same figure we also show a molecular orbital 
calculation (dashed line) of the anisotropy due to a 
hole in the $\rm Cu 3d_{x^2-y^2}$ state corresponding 
to Cu-O layers in undoped insulating $\rm RBa_2Cu_3O_6$.
Again we are limited to a qualitative approach because 
LDA predicts an incorrect metallic ground-state for
$\rm RBa_2Cu_3O_6$ with anisotropy similar to that for
$\rm RBa_2Cu_3O_{7-\delta}$.
The calculation accounting for disorder mimics the strong 
suppression of the first peak in the measured anisotropy 
for $\rm PrBa_2Cu_3O_{7-\delta}$ which also tends 
towards the behaviour calculated for undoped Cu-O layers. 
We argue below how $\rm Pr_{Ba}$ disorder could in fact lead
to a situation where carriers are not doped in to 
the Cu-O layers, effectively killing both metallic
behaviour and superconductivity in 
$\rm PrBa_2Cu_3O_{7-\delta}$.

The charge-transfer model for the effect of oxygen
doping in $\rm YBa_2Cu_3O_{7-\delta}$ \cite{cava,jorgensen}
is well established, and describes the mechanism 
of transfer of positive charge from the Cu-O chains
to the Cu-O layers as oxygen is added to the chains.
Two possible mechanisms may exist to block or modify this
charge transfer. The first that we propose,
supposes that if Pr is present at the Ba site, the needed 
charge is supplied by $\rm Pr_{Ba}$ (possibly in a 4+ 
oxidation state because the chemical environment on this site 
is different), inhibiting hole creation in the layers.
However, the chains would maintain their microscopic metallic 
character as has been experimentally observed and mentioned above.
It is significant that according to Cava et al. \cite{cava} the 
total charge transfer associated with $\rm T_c$
going from 0K to 90K is 0.08{\em e}/Cu-O layer in the unit cell. 
Knowing that the 
concentration of $\rm Pr_{Ba}$ defects is a few percent, we
have here a hole-depleting mechanism which is of the
same magnitude as the hole-creating mechanism of charge transfer.
Experimental observations of hole-depletion in the 
Cu-O layers include optical 
conductivity \cite{takenaka}, NMR 
\cite{reyes}, Hall-effect \cite{matsuda} 
and transport \cite{neumeier} measurements.
Most experimental evidence favours a valence for Pr close 
to +3 (compatible, in our model, 
with Pr on the rare-earth site) but also indicates the probable
presence of a small fraction of +4 valent Pr 
\cite{booth,jostarndt,neumeier,guillaume},
which can be explained if $\rm Pr_{Ba}$ had a valence of +4.
Studies on the influence of ion-size in 
$\rm R_{1-x}Pr_{x}Ba_2Cu_3O_{7-\delta}$ show that at constant Pr 
concentration, $\rm T_c$ decreases while 
the antiferromagnetic ordering temperature of Pr ions, $\rm T_N$, 
increases with the ionic radius of R \cite{guan}.
It is reasonable to assume that
the larger the ionic radius of the host rare earth R, 
the easier its substitution by Pr,
leading also to the higher $\rm T_N$.
Lastly, Zou {\em et al.} \cite{zou} found an anomalously
short c lattice parameter in non-superconducting 
$\rm PrBa_2Cu_3O_{7-\delta}$, which could be due to the 
smaller size of the $\rm Pr^{4^+}$ ion.

We have further analyzed the anisotropies shown in 
Fig.\ \ref{fig1} and Fig.\ \ref{fig2} by computing the
power spectral density:
$\left|\int_{-\infty}^{\infty}\left(J_{100/010}(p)-J_{110}(p)
\right)\exp\left\{{-\imath p r}\right\}dp\;\right|^2$.
Compton profiles, and hence anisotropies, are modulated 
with oscillations corresponding to characteristic distances 
over which wavefunctions are coherent in 
given crystallographic directions \cite{malcolm,pattison}. 
Peaks in the power spectral density then indicate 
these characteristic distance scales
as shown in  Fig.\ \ref{fig3}. Only the positions and relative
intensities of the peaks in each spectrum are significant 
since the theoretical anisotropies have been scaled.
The top panel shows that experiment 
(solid line with dots) and theory (solid line) 
for $\rm YBa_2Cu_3O_{7-\delta}$ as well as theory for
$\rm PrBa_2Cu_3O_{7-\delta}$ (dashed line) are in
agreement and that the charge carriers are of a delocalized
nature. The periodic structure of 
the anisotropy shown in Fig.\ \ref{fig1} and the corresponding
peaks in the power spectral density in Fig.\ \ref{fig3} are
directly related to 
the phases of the wavefunction associated with the four O atoms 
surrounding a given Cu atom in the Cu-O layers \cite{barplatz}.
In the bottom panel the experiment
for $\rm PrBa_2Cu_3O_{7-\delta}$ (solid line with dots) 
shows both a tendency for strong suppression 
of the longer range coherence 
peaks as in the disordered model (solid line), as well as a
shifting of the remaining coherence peak to shorter length scales
similar to the simple model of insulating, undoped
Cu-O planes (dashed line). This is a clear indication of hole
wavefunction localization in the Cu-O layers and of the disorder
on the Pr and Ba sites.
On-going experiments in under-doped $\rm YBa_2Cu_3O_{7-\delta}$ 
confirm the possibility to monitor wavefunction coherence length
behaviour near the metal-insulator transition 
using Compton profiles anisotropies.

With these arguments in mind we can state that superconductivity
in $\rm PrBa_2Cu_3O_{7-\delta}$ appears with decreasing 
$\rm Pr_{Ba}$ defects and that this is also the reason for the 
scatter in the measured $\rm T_c$ values which correspond to 
varying degrees of such disorder, dependent
on the method and particular conditions of sample fabrication.
Mazin \cite{liechmazin} has independently 
come to a similar conclusion
arguing however that disorder effectively localizes holes
in the Cu-O layers in $\rm PrBa_2Cu_3O_{7-\delta}$ which are
of a different nature with respect to $\rm YBa_2Cu_3O_{7-\delta}$
due to hybridization. This possible
second mechanism for hole depletion would also lead to a 
metal-insulator transition in the Cu-O layers and therefore
to similar results for the anisotropy.
One or both of these mechanisms
may be active and we point out that disorder-induced 
disruption of charge transfer would provoke insulating
behaviour independently of the precise nature of 
the holes which would have been created 
in the absence of disorder.
Further support for this picture is provided by the fact 
that by using a combined strategy of 
minimizing substitution disorder by quenching 
the melt from high temperatures after crystal growth 
{\em and} avoiding tetravalent $\rm Pr_{Ba}$, by 
using very low oxygen partial pressures during growth, we have
succeeded in obtaining small flux-grown crystallites of 
$\rm PrBa_2Cu_3O_{7-\delta}$ with a
$\rm T_c$ of 90 K and a bulk Meissner effect \cite{erb2,erb3}.
The oxygenation used for these samples was the same as for
90K superconducting $\rm YBa_2Cu_3O_{7-\delta}$
implying similar, optimal doping in both.
We expect the electron momentum density 
in superconducting $\rm PrBa_2Cu_3O_{7-\delta}$ and
$\rm YBa_2Cu_3O_{7-\delta}$ to be similar, but at
the moment such samples are much too small for Compton
profile measurements.

In conclusion we provide a consistent explanation of the 
observed behaviour in $\rm PrBa_2Cu_3O_{7-\delta}$. 
Our data explicitly shows a strong 
suppression of the wavefunction coherence
length in $\rm PrBa_2Cu_3O_{7-\delta}$ 
due to disorder in the Ba and Pr planes and 
the metal-insulator transition of the Cu-O layers.
The presence of some Pr on the Ba site, probably in a 4+ 
oxidation state, while maintaining the carrier 
concentration in the Cu-O chains, inhibits hole-doping 
in the Cu-O planes and/or localizes doped carriers. 
We further propose that superconductivity appears if this 
substitution is suppressed and our preliminary experiments
yielding superconducting $\rm PrBa_2Cu_3O_{7-\delta}$
strongly support this model.

We acknowledge help from P. Fajardo, D. Vasumathi and B. Revaz,
and discussions with S. Massidda, S. Ishibashi and T. Jarlborg.

\begin{figure}
\caption{Anisotropy ($\rm J_{[100/010]}-J_{[110]}$) in the 
electron momentum density of $\rm YBa_2Cu_3O_{7-\delta}$. 
Note the good agreement between the experiment and
the LDA calculations ($\rm YBa_2Cu_3O_{7-\delta}$: solid line
$\rm PrBa_2Cu_3O_{7-\delta}$: dashed line).
Calculations are scaled to the experimental variation.
}
\label{fig1}
\end{figure}
\begin{figure}
\caption{Anisotropy ($\rm J_{[100/010]}-J_{[110]}$) in the 
electron momentum density of $\rm PrBa_2Cu_3O_{7-\delta}$. 
The experiment differs remarkably from the LDA calculation
(Fig.\ \ref{fig1}). Two simple theoretical models accounting for
$\rm Pr_{Ba}$ disorder (solid line) and undoped Cu-O planes
(dashed line) are also shown.
}
\label{fig2}
\end{figure}
\begin{figure}
\caption{
Power spectral densities of anisotropies. 
The peaks indicate characteristic 
distance scales over which 
wavefunctions are coherent.
Top panel: Experiment (solid line with dots)
and theory(solid line) for $\rm YBa_2Cu_3O_{7-\delta};$
Theory for $\rm PrBa_2Cu_3O_{7-\delta}$ (dashed line).
Bottom panel: In $\rm PrBa_2Cu_3O_{7-\delta}$,
experiment (solid line with dots) shows a 
strong suppression of the longer range coherence 
peaks as in the disordered model (solid line) as well as a
shifting of the remaining coherence peak to shorter length scales
similar to the model for insulating, undoped
Cu-O layers (dashed line).
}
\label{fig3}
\end{figure}

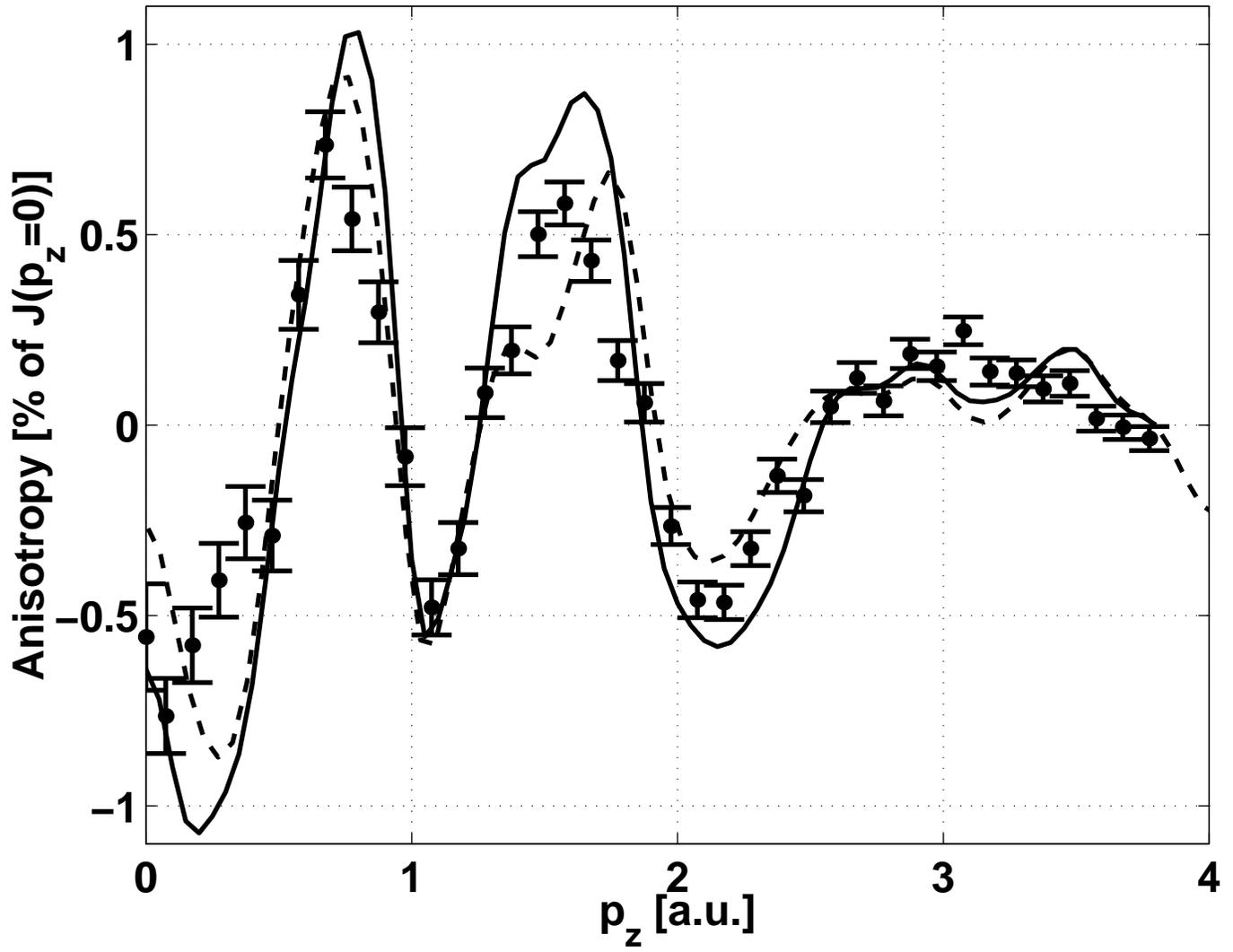
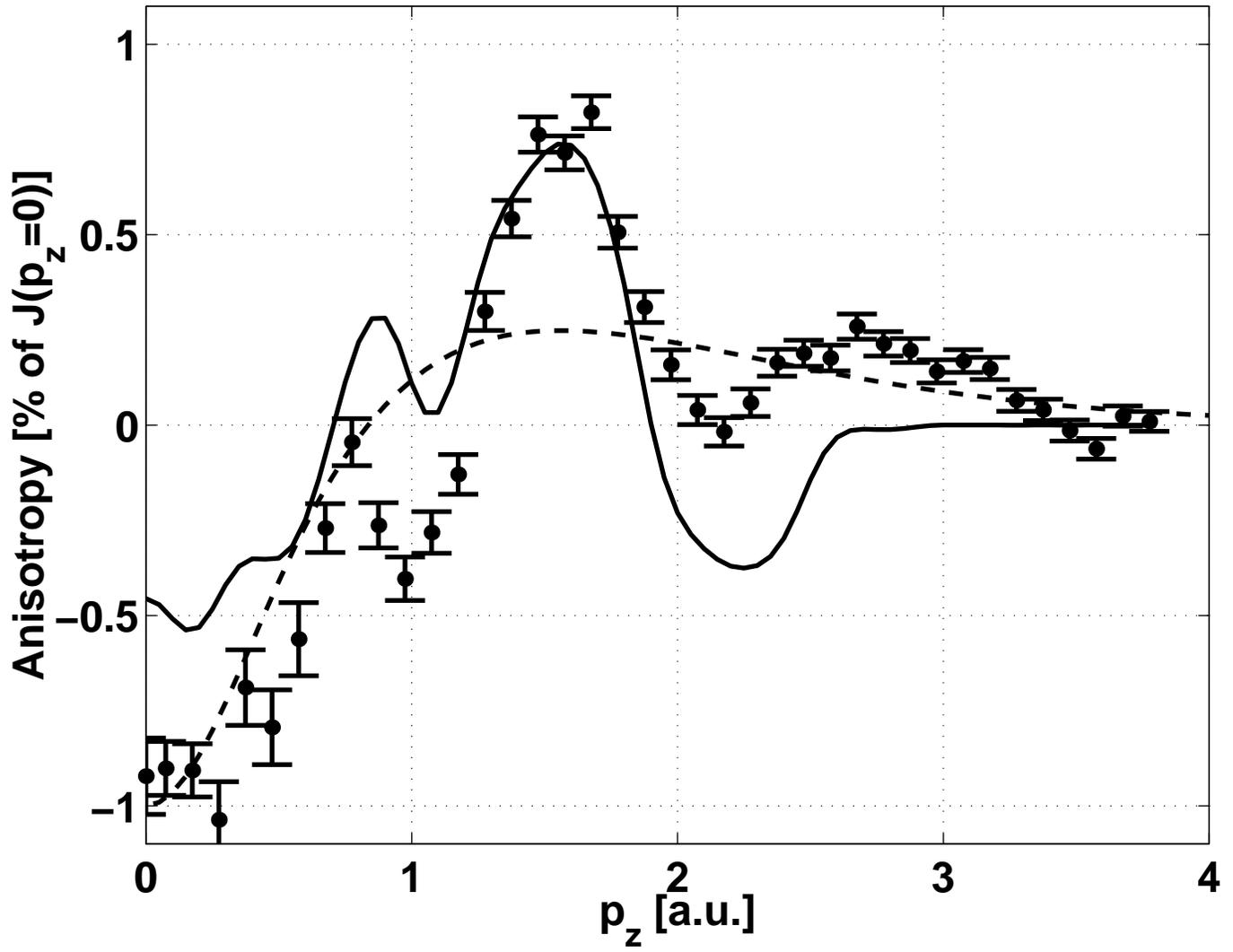
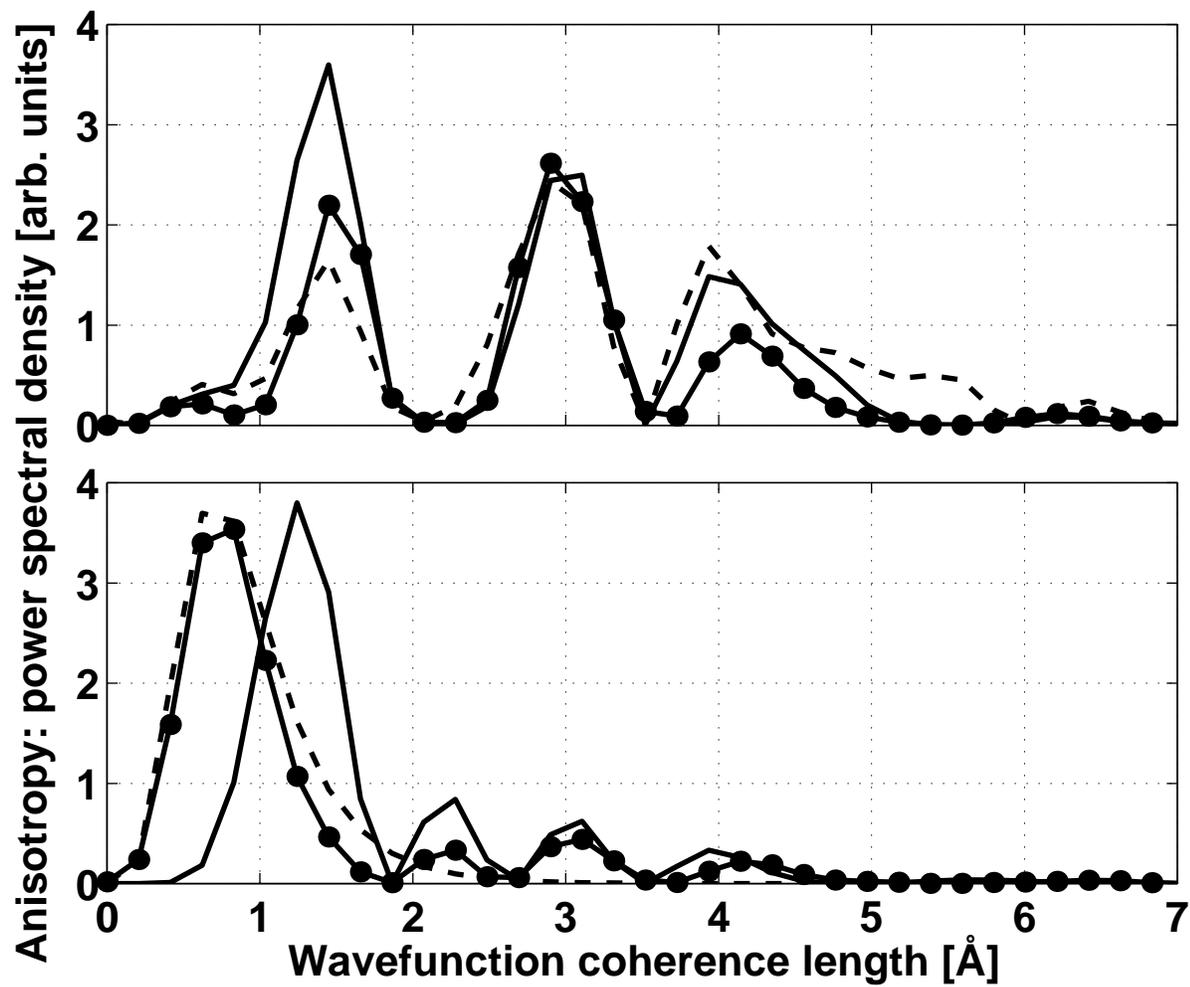
\end{document}